\documentclass[12pt,preprint]{aastex}

\newcommand{\bq}{\begin{equation}}
\newcommand{\eq}{\end{equation}}

\begin{document}
 
\title{Are X-ray Clusters Cooled by Heat Conduction to the Surrounding
Intergalactic Medium?}

\author{Abraham Loeb} 
\affil{Astronomy Department, Harvard University, 60
Garden St., Cambridge, MA 02138;\\ 
aloeb@cfa.harvard.edu}

\begin{abstract}
We show that X-ray clusters would have cooled substantially over a Hubble
time by transport of heat from their hot interior to the their envelope, if
the heat conductivity had not been heavily suppressed relative to the
Spitzer value due to magnetic fields.  The suppression is required in order
for the observed abundance of hot X-ray clusters to be consistent with
predictions from popular cosmological models.  If a similar or stronger
suppression factor applies to cluster cores, then thermal conduction can
not be the mechanism that prevents cooling flows there.

\end{abstract}

Key Words: galaxies: clusters: general -- cooling flows -- conduction --
hydrodynamics 

\vspace{-.1in}
PACS: 95.30.Tg, 95.30.Lz

\section{ Introduction }

Recently, the old idea that heat conduction may suppress cooling flows in
X-ray clusters (Binney \& Cowie 1981; Tucker \& Rosner 1983; Bertschinger
\& Meiksin 1986; Bregman \& David 1988; Gaetz 1989; Rosner \& Tucker 1989;
David et al. 1992; Pistinner \& Shaviv 1996; Dos Santos 2001) has been
revived due to the apparent lack of strong cooling flows in {\it Chandra}
and {\it XMM-Newton} data (Fabian et al. 2001; B{\" o}hringer et al. 2001;
Molendi \& Pizolato 2001). It was argued that if the heat conductivity is
not suppressed by more than a factor of a few relative to the Spitzer
value, then the inward heat flow due to the positive temperature gradient
in cluster cores would be sufficient to compensate for the energy loss
caused by Bremsstrahlung cooling of the gas (Narayan \& Medvedev 2001;
Gruzinov 2002; Voigt et al. 2002).

The purpose of this paper is to caution that a large heat conduction
coefficient would also lead to dramatic {\it cooling} of the entire cluster
gas due to energy transport {\it outwards} into the cooler, surrounding
intergalactic medium.  By analyzing {\it ASCA} data for 30 X-ray clusters,
Markevitch et al. (1998) have identified a composite temperature profile
that declines by a factor of $\sim 2$ from the core out to about half the
virial radius (see also Finoguenov et al. 2001). This profile has been
confirmed in a recent {\it BeppoSAX} data set of 21 clusters (De Grandi \&
Molendi 2002).  Here we use this profile to calculate the resulting
conductive heat transfer from the cluster interior to the surrounding
envelope.

We note that the above {\it ASCA} results were challenged in several
clusters by preliminary {\it XMM-Newton} data which showed a nearly
isothermal profile (Arnaud et al. 2001a,b; Pratt et al. 2001).  However,
since clusters represent the hottest structures in the universe there is no
doubt that they are surrounded by cooler gas at a sufficiently large
radius.  In fact, if the temperature profile is nearly flat all the way out
to the radius where the surface brightness of the cluster is too faint to
be detectable, then the temperature gradient around the (more distant)
virial radius of the cluster must be even steeper than assumed here (since
the temperature must eventually approach the ambient value over a shorter
range of radii).  Hence, the heat flux that we calculate based on the
Markevitch et al. (1998) temperature profile provides a conservative {\it
lower limit} for the conductive energy loss of cluster interiors.

Infall of gas onto the cluster could in principle suppress conduction of
heat outwards. However, numerical simulation show that much of this infall
is confined within filaments that cover only a small fraction of the
surface area of the accreting object, and the gas flow is often clumpy and
episodic (e.g., Yoshida et al. 2002, and references therein). Hence,
conduction is unlikely to be suppressed by infall in all directions at all
times. In this paper we focus our attention on the virialized region of the
cluster, where the gas is stationary and in hydrostatic equilibrium.  Since
most of the mass is located in the outer envelope of the cluster, a
considerable transport of heat out of the cluster core would only have a
minor effect on the cluster envelope.  For example, a reduction by $\sim
50\%$ in the X-ray temperature of the cluster -- which is dictated by the
core at $\sim 300~{\rm kpc}$ -- would typically only lead to a $\sim
5$--$10\%$ increase in the outer envelope temperature at a radius of $\sim
3~{\rm Mpc}$.  In the following we examine the question of whether the
cores of clusters, which dominate their X-ray emission, would cool through
thermal conduction by more than they are allowed to cool over a Hubble
time, given the observed temperature profiles of Markevitch et al. (1998).
Throughout the discussion we assume a fully-ionized, hydrogen-helium plasma
with a helium mass fraction of $24\%$, for which the mean particle mass
(including the electrons) is $\mu=0.59$ in units of the proton mass $m_p$.

\section{ Cooling of X-ray Clusters by Heat Conduction}

The time over which thermal energy is drained out of an X-ray cluster by
conduction can be found by dividing the total thermal energy of the
cluster, $(M_g/\mu m_p)({3\over 2} k_B {\bar T})$, by the rate of
conductive heat loss across its boundary radius $r_b$,
\begin{equation}
t_{\rm cond}= {{3\over 2} (M_g/\mu m_p) k_B {\bar T}\over 4\pi r_b^2
\left[\kappa \vert\partial T/\partial r\vert\right]_{r_b}} ,
\label{eq:t_cond}
\end{equation}
where ${\bar T}=M_g^{-1} \int_0^{r_b} T(r) \rho_g(r) 4\pi r^2 dr$ is the
(mass-weighted) mean temperature of the cluster, $\rho_g(r)$ is the mass
density of the gas, $M_g=\int_0^{r_b}\rho_g(r)~ 4\pi r^2 dr$ is the total
gas mass out to the radius $r_b$, $\kappa$ is the coefficient of heat
conductivity, and $k_B$ is Boltzmann's constant.  The hydrostatic
equilibrium equation, $(GM_{\rm tot}\rho_g/r^2) =- \partial_r (\rho_g k_B
T/\mu m_p)$, yields
\begin{equation}
{M_g\over r_b}= - f_g \left[\left({k_B
T\over G \mu m_p}\right) \left({\partial \ln\rho_g\over \partial\ln r} +
{\partial \ln T\over \partial\ln r} \right)\right]_{r_b},
\label{eq:Mr}
\end{equation}
where $f_g=(M_g/M_{\rm tot})$ is the gas mass fraction at $r_b$, $G$ is
Newton's constant, and $\partial_r\equiv {\partial\over \partial r}$.  By
substituting equation~(\ref{eq:Mr}) into equation~(\ref{eq:t_cond}) we get
\begin{equation}
t_{\rm cond}= {3\over 8\pi} {f_g k_B {\bar T}\over G \mu^2 m_p^2} \left[
\kappa^{-1}\left(1+ {\partial_r \ln \rho_g \over \partial_r \ln T
}\right)\right]_{r_b} .
\label{eq:t-beta}
\end{equation}
Thus, the characteristic cooling time for the entire cluster depends on the
boundary values of the conductivity coefficient, $\kappa(r_b)$, and the
effective ``adiabatic index'' of the gas, $\gamma_{\rm eff} \equiv [1+
(\partial_r \ln T / \partial_r \ln \rho_g)]_{r_b}$.

As customary in the literature, we identify the boundary of the hot
interior of the cluster as the radius, $r_{180}$, where the average
interior density of the gas is $180$ times the mean cosmological density.
The universal temperature profile derived by Markevitch et al. (1998) from
ASCA data implies $[\partial \ln T/\partial \ln r]_{r_b} \approx -0.4$,
$\beta_b\equiv [T(r_b)/ {\bar T}]\approx 0.6$, and $\gamma_{\rm eff}\approx
1.24$.
In order to get a numerical value for the cooling time $t_{\rm cond}$, we
normalize the conductivity coefficient by the Spitzer (1962) value,
$\kappa_{\rm Sp}(r_b)=5 \times 10^{29} k_B (\beta_b {\bar
T}_{10})^{5/2}~{\rm cm^{-1}~s^{-1}}$ where ${\bar T}_{10}=(k_B{\bar
T}/10~{\rm keV})$, and define the suppression fractor $\eta\equiv
\kappa/\kappa_{\rm Sp}$. By substituting the above parameter values into
equation~(\ref{eq:t-beta}) we get\footnote{Note that the heat flux
saturates when the Coulomb mean-free-path $\lambda_{\rm Coul}$ becomes
longer than the characteristic scale of the temperature variation
$(\partial \ln T/\partial r)^{-1}$ (Sarazin 1988). For a conductivity
coefficient which is lower than the Spitzer value by a factor of $\eta$,
the effective mean-free-path (at the same thermal speed) is $\lambda_{\rm
eff}\sim \eta \lambda_{\rm Coul}$. From the $T$-vs-$r_{180}$ relation of
Evrard et al. (1996) we get $(\lambda_{\rm eff}/r_{180})\sim 0.1 (\eta/0.1)
T_{10}^{3/2}$.},
\begin{equation}
t_{\rm cond}= 5\eta^{-1}~\left({f_g\over 0.15}\right) 
\left({k_B {\bar T}\over 10~{\rm keV}}\right)^{-3/2}~{\rm Gyr}~.
\label{eq:t-value}
\end{equation}
Over a cluster lifetime, $\tau_{\rm cl}$, the cluster temperature is
expected to decline by a fraction
\begin{equation}
{\Delta {\bar T}\over {\bar T}} \approx {\tau_{\rm cl}\over t_{\rm cond}}=
0.2 \left({\eta\over 0.1}\right) \left(\tau_{\rm cl}\over 10~{\rm
Gyr}\right) \left({f_g\over 0.15}\right)^{-1} \left({k_B {\bar T}\over
10~{\rm keV}}\right)^{3/2} ,
\label{eq:T-change}
\end{equation}
where we have assumed that $\tau_{\rm cl}\ll t_{\rm cond}$.  Note that the
cluster lifetime cannot be shorter than the sound crossing time across its
diameter, $\sim 6 (r_{180}/5~{\rm Mpc}){\bar T}_{10}^{-1/2}~{\rm Gyr}$.
Cooling is inevitable (and cannot be eliminated by adiabatic heating of the
gas as it settles towards the center) since the gravitational potential
that confines the gas is dominated by the dark matter.

Equation~(\ref{eq:T-change}) implies that hot X-ray clusters in the local
universe must have been even hotter when they formed.  The abundance of
X-ray clusters is exponentially suppressed at high temperatures (Henry \&
Arnaud 1991; Eke et al. 1996; Viana \& Liddle 1996; Pierpaoli et al. 2001;
Ikebe et al. 2002) and even more so at earlier cosmic times (Fan, Bahcall,
\& Cen 1997; Bahcall \& Fan 1998; Evrard et al. 2002).  It therefore
becomes exponentially more difficult to account for the abundance of hot
clusters in the present-day universe if those same clusters had to be even
hotter $\sim 5$ billion years ago.  The agreement between the measured
abundances of hot ($k_B {\bar T} >6~{\rm keV}$) clusters and those
predicted by popular cosmological models (see Figures 9 and 12 in Evrard et
al. 2002) would be significantly spoiled unless we require $(\Delta{\bar
T}/ {\bar T})\la 0.3$. From equation~(\ref{eq:T-change}) we then find
$\eta\la 0.15 (\tau_{\rm cl}/10~{\rm Gyr})^{-1} {\bar T}_{10}^{-3/2}$. Note
that this constraint is strongest for the hottest clusters where the total
thermal energy is $\sim 10^{64}$ ergs, and in which plausible astrophysical
heating sources (such as supernovae or active galactic nuclei) are unable
to compensate for the energy loss due to thermal conduction. Of particular
interest is the cluster 1E 0657-56 at $z=0.296$ for which the inferred
emissivity-weighted temperature is $14.8^{+1.7}_{-1.2}$ keV, although it
may be a relatively young merger (Tucker et al. 1998; Markevitch et al.
2002).  Since heat conduction is temperature dependent, it distorts the
shape of the abundance distribution of cluster temperatures in a generic
way that is not degenerate with variations in cosmological parameters such
as the normalization of the power-spectrum of density
fluctuations. Inclusion of thermal conduction in future high-resolution
hydrodynamic simulations can be used to refine the above upper limit on
$\eta$. In these simulations, it would be important to treat the electron
and ion temperatures separately since the thermal equilibiration 
timescale is long near the virialization shock of 
clusters (Fox \& Loeb 1997).

We reiterate that the emission-weighted temperature of an X-ray cluster is
dominated by its core which amounts to only a small fraction of the total
cluster mass. A substantial reduction in the core temperature would merely
result from the transfer of heat out of the hot core to the cooler,
stationary cluster envelope without requiring that this heat be transferred
further to accreting material in the surrounding large-scale structure.
Nevertheless, heat conduction is inevitably expected to preheat the
surrounding gas and suppress smooth accretion of gas onto the gravitational
potential well of the hottest clusters. If substantial, such a suppression
would have implied that the baryon mass fraction in the hottest clusters is
lower than its cosmic average, a result that would have been at odds with
the concordance model of Big Bang Nucleosynthesis and structure formation
(Burles et al. 2001; Tegmark, Zaldarriaga, \& Hamilton 2002).

\section{Conclusion}

We find that heat conduction must be suppressed by a factor $\eta\la 0.15
(\tau_{\rm cl}/10~{\rm Gyr})^{-1} {\bar T}_{10}^{-3/2}$ relative to the
Spitzer value, or else the cores of X-ray clusters would have cooled
significantly over their lifetime.  This suppression can be naturally
produced (see Malyshkin \& Kulsrud 2000 and \S 5 in Malyshkin 2001) by the
magnetic fields that are inferred to exist in the halos of X-ray clusters
(Carilli \& Taylor 2002).  The recent {\it Chandra} detections of sharp
temperature jumps (cold fronts) in several clusters (Markevitch et
al. 2000; Vikhlinin et al. 2001a) indicate an even stronger suppression of
heat conduction across these jumps (Markevitch et al. 2000; Ettori \&
Fabian 2000).  Similarly, a suppression factor of $\eta \sim 1$--$3\%$ is
required to account for the conditions in the interstellar gas of cluster
galaxies (Vikhlinin et al. 2001b). If this suppression applies also to the
diffuse gas in cluster cores, then thermal conduction could not account for
the apparent lack of cooling flows in them. Mixing of gas due to mergers or
central heating sources, such as active galactic nuclei, could in principle
compensate for the radiative losses in these environments (B{\" o}hringer
et al. 2002; Churazov et al. 2002). In this paper we have shown that if the
thermal conductivity had been comparable to the Spitzer value as recently
suggested (Voigt et al. 2002), then cooling flows would have not been
suppressed but rather {\it induced} in hot X-ray clusters.

\acknowledgements

I thank Maxim Markevitch and Jerry Ostriker for useful comments on the
manuscript.  This work was supported in part by the grants 5-7768 from NASA
and AST-9900877, AST-0071019 from NSF.

\end{document}